\documentclass[10pt]{article}
\usepackage{graphicx}

\def\Title#1{\begin{center} {\Large #1 } \end{center}}
\def\Author#1{\begin{center}{ \sc #1} \end{center}}
\def\Address#1{\begin{center}{ \it #1} \end{center}}

\newcommand\pubblock{\rightline{\begin{tabular}{l} Proceedings of the Fifth Annual LHCP\\ \pubnumber\\
         \pubdate  \end{tabular}}}

\newenvironment{Abstract}{\begin{quotation} \begin{center} 
             \large ABSTRACT \end{center}\bigskip 
      \begin{large}}{\end{large} \end{quotation}}

\newenvironment{Presented}{\begin{quotation} \begin{center} 
             PRESENTED AT\end{center}\bigskip 
      \begin{center}\begin{large}}{\end{large}\end{center} \end{quotation}}

\def\beq{\begin{equation}}
\def\eeq#1{\label{#1}\end{equation}}
\def\eeqn{\end{equation}}

\def\beqa{\begin{eqnarray}}
\def\eeqa#1{\label{#1}\end{eqnarray}}
\def\eeqan{\end{eqnarray}}

\let\bar=\overbar

\def\Dslash{\not{\hbox{\kern-4pt $D$}}}
\def\dslash{\not{\hbox{\kern-2pt $\del$}}}

\def\msb{{\bar{\ssstyle M \kern -1pt S}}}

\newcommand*{\pT}{\ensuremath{p_{\text{T}}}\xspace}

\newcommand*{\GeV}{\ensuremath{\text{Ge\kern -0.1em V}}}
\newcommand*{\TeV}{\ensuremath{\text{Te\kern -0.1em V}}}

\usepackage{color}
\usepackage{amsmath}
\usepackage{xspace}
\usepackage{subfig}
\usepackage{lineno}

\newcommand\pubnumber{ ATL-PHYS-PROC-2017-089 }

\newcommand\pubdate{\today}

\def\affiliation{
On behalf of the ATLAS Collaboration\\
The University of British Columbia -- TRIUMF}

\def\support{\footnote{The author acknowledges support from the Vanier Canada Graduate Scholarship program, and the Natural Sciences and Engineering Research Council of Canada.}}

\begin{document}

\large
\begin{titlepage}
\pubblock

\vfill
\Title{ Searching for New High Mass Phenomena Decaying to Muon Pairs using Proton-Proton Collisions at $\lowercase{\sqrt{s}} = 13~\TeV{}$ with the ATLAS Detector at the LHC }
\vfill

\Author{ S\'ebastien Rettie \support }
\Address{\affiliation}
\vfill
\begin{Abstract}

We present a search for new high mass phenomena using the latest data collected by the ATLAS detector at the LHC, corresponding to 36.1 fb$^{-1}$ at $\sqrt{s}=13~\TeV{}$. The search is conducted for both resonant and non-resonant new phenomena in dimuon final states. The dimuon invariant mass spectrum is the discriminating variable used in the search. No significant deviations from the Standard Model expectation are observed. Lower limits are set on the signal parameters of interest at 95\% credibility level, using a Bayesian interpretation. In particular, a Sequential Standard Model Z' resonance is excluded for masses below $4.0~\TeV{}$.

\end{Abstract}
\vfill

\begin{Presented}
The Fifth Annual Conference\\
 on Large Hadron Collider Physics \\
Shanghai Jiao Tong University, Shanghai, China\\ 
May 15-20, 2017
\end{Presented}
\vfill
\end{titlepage}
\def\thefootnote{\fnsymbol{footnote}}
\setcounter{footnote}{0}
%

\normalsize 


\section{Introduction}

The Standard Model (SM) of particle physics is a very successful predictive theory which explains the fundamental interactions of elementary particles in the universe, except for gravity. However, the SM is known to be an effective theory that is valid only in a low energy regime, called the electroweak scale, and does not account for many observed experimental results. For example, it does not offer a satisfying explanation for neutrino masses or dark matter. Hence, it is clear that to fully understand and explain nature, a theoretical framework that goes beyond the Standard Model (BSM) is required. While high mass resonances do not offer a complete solution to the problems mentioned above, many BSM theories predict their existence. For example, extra dimensional models \cite{Randall:1999ee} and new gauge boson models \cite{Langacker:2008yv} both have the common goal of reconciling the very different scales of electroweak symmetry breaking and high mass scales, and predict the existence of high mass resonances. Thus, finding high mass resonances would help validate these theories, which do offer solutions to the aforementioned problems. The ATLAS experiment \cite{Aad:2008zzm} at the Large Hadron Collider (LHC) has collected 36.1 fb$^{-1}$ of data at $\sqrt{s}=13~\TeV{}$. Using this dataset, we have searched for new high mass phenomena with dimuon final states. Both resonant and non-resonant phenomena are considered.

\section{Event Selection}

In order to select events with two muons in the final state, a series of cuts is applied to each event. At the event level, each event is required to be in the so-called Good Run List (GRL) as defined by the ATLAS data quality group. An event is contained in the GRL if it occured during a period where the proton beams in the LHC were stable and all relevant detector systems were functional. Next, each event is required to pass a single-muon trigger which requires at least one isolated muon with transverse momentum, \pT, greater than 26~\GeV{}, or one muon with \pT greater than 50~\GeV{}. Event cleaning is then applied, where events are rejected if they are flagged as incomplete. An incomplete event is one where noise bursts or data corruption occurs in the electromagnetic or hadronic calorimeter. Finally, each event is required to contain at least two combined muons. A combined muon is one where the track reconstruction for the muon candidate is performed independently in the Inner Detector (ID) and the Muon Spectrometer (MS), and a combined track fit is formed with a global refit that uses hits from both the ID and MS subdetectors. 

Each selected muon candidate must have a track consistent with the primary vertex both along the beamline and in the transverse plane, have \pT greater than 30~\GeV{}, and satisfy the high-\pT working point described in~\cite{Aad:2016jkr}. One of the most important criteria of the high-\pT working point is that each muon is required to have hits in 3 separate muon stations in the MS. Finally, muons are also required to fulfil track-based isolation requirements in order to reduce the background from hadron decays inside jets.

\begin{figure}[!h]
\begin{center}
\subfloat[][]{
\includegraphics[width=0.48\linewidth]{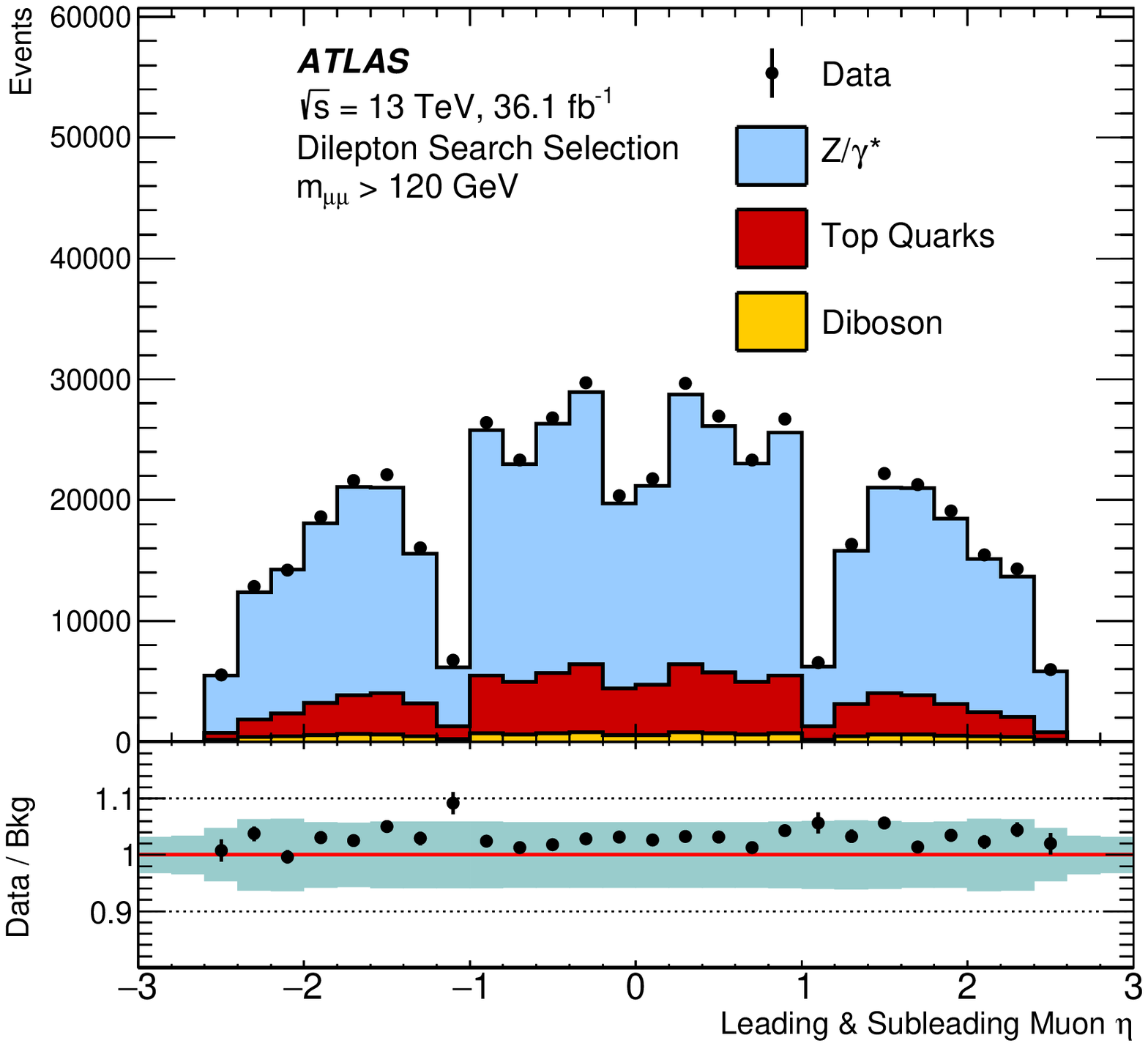}\label{subfig:EtaDist}
}
\subfloat[][]{
\includegraphics[width=0.48\linewidth]{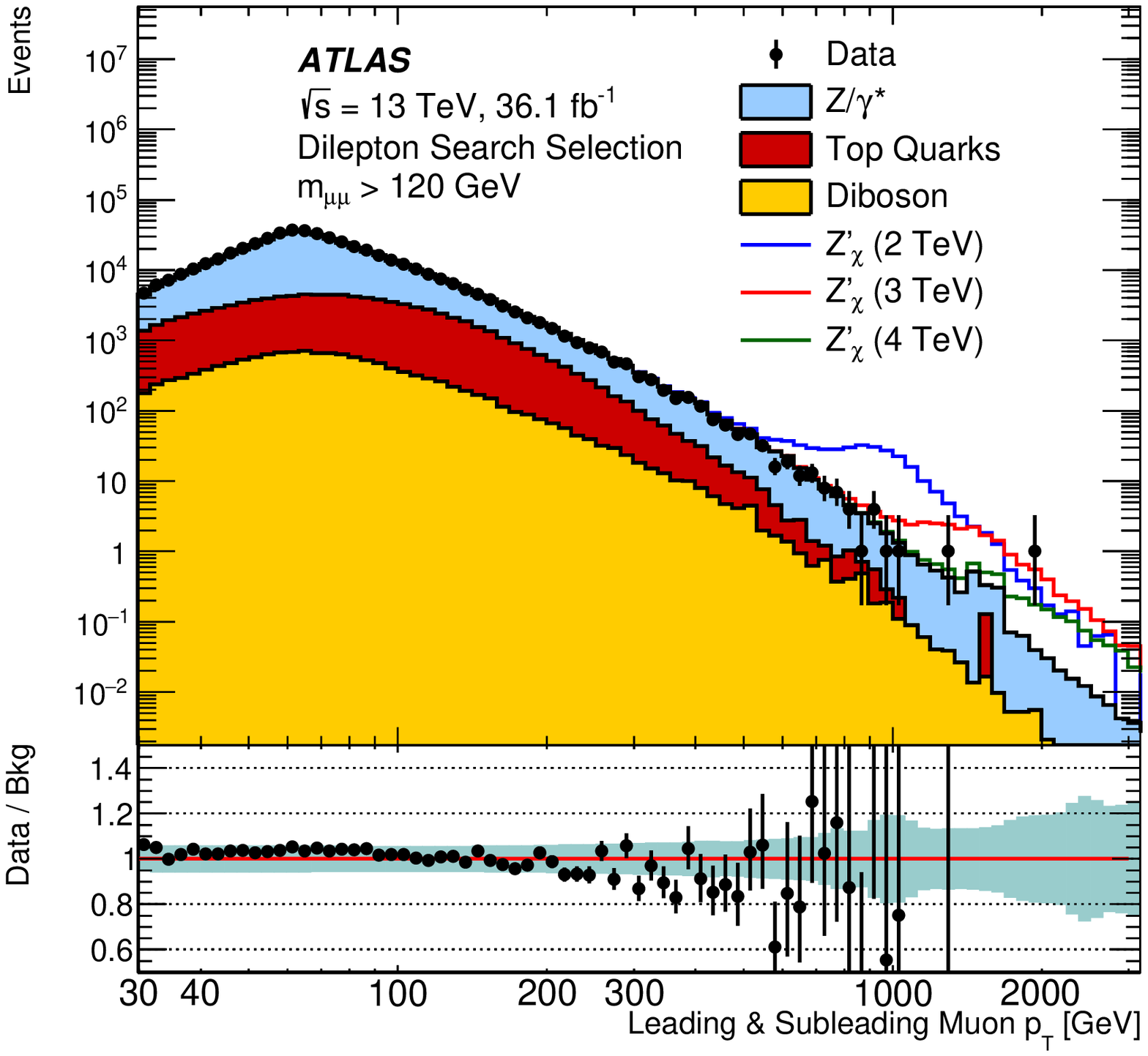}
\label{subfig:pTDist}
}
\end{center}
\caption{\protect\subref{subfig:EtaDist} $\eta$ and \protect\subref{subfig:pTDist} \pT distributions for leading \& subleading muons in events with $m_{\mu\mu} > 120~\GeV{}$ \cite{Aaboud:2017buh}.}
\label{fig:Kinematics}
\end{figure}

\begin{figure}[!h]
\centering
\includegraphics[width=0.5\linewidth]{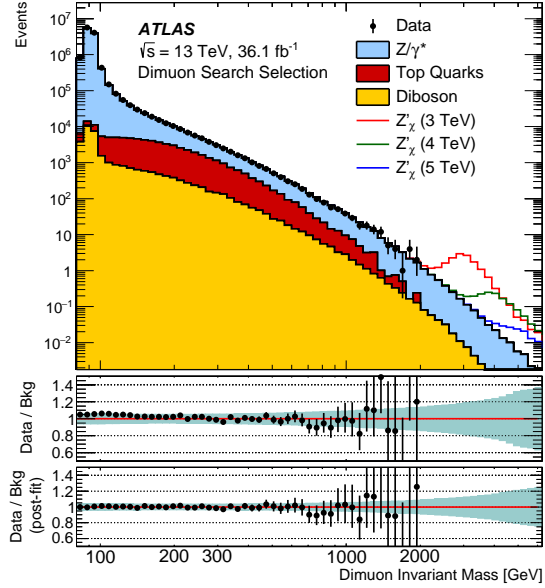}
\caption{Dimuon invariant mass distribution after event selection \cite{Aaboud:2017buh}.}
\label{fig:InvMass}
\end{figure}

The invariant mass of a dimuon pair, $m_{\mu\mu}$, is chosen as the discriminating variable for this search. It is calculated using the muon pair in the event having the highest scalar sum \pT. In addition, the pair must have opposite sign charge, and have $m_{\mu\mu} > 80~\GeV{}$. The kinematic distributions of the leading and subleading muons are shown in Figure \ref{fig:Kinematics}. The invariant mass distribution of the dimuon system is shown in Figure \ref{fig:InvMass}.



\section{Background Processes}

The background processes for this search are evaluated using Monte Carlo event generators and the full simulation of the ATLAS detector based on GEANT4 \cite{AGOSTINELLI2003250}. The three main background contributions are Drell-Yan (DY) production, top quark production, and diboson (WW, WZ, ZZ) production. A summary of the various properties of the background processes considered is presented in Table \ref{tab:Backgrounds}. The event generators used are Powheg Box \cite{Alioli:2010xd} and Sherpa \cite{Gleisberg:2008ta}, and the event showering is done using either Pythia \cite{Sjostrand:2007gs} or Sherpa. The background component due to QCD is negligible when considering the dimuon final state.

The theoretical systematic uncertainties pertaining to the parton distribution function (PDF) of the incoming partons include the choice of the PDF used, the variations of the PDF eigenvector sets, and the variations in the PDF renormalisation and factorisation scales. Other theoretical systematic uncertainties include the variations in the $\alpha_s$ coupling constant, the differences between the additive and factored treatment of the electroweak corrections (the additive treatment is used as the nominal treatment in this search), and photon-induced corrections. The dominant theoretical systematic uncertainty in this analysis is the DY background PDF eigenvector set variation, resulting in a relative systematic uncertainty of $\sim 8\%$ ($\sim 13\%$) at $m_{\mu\mu} = 2~\TeV{}$ ($m_{\mu\mu} = 4~\TeV{}$). The experimental systematic uncertainties include the beam energy uncertainty, trigger efficiency uncertainty, and muon energy scale and resolution uncertainty. In addition, muon idendification, reconstruction, and isolation efficiency uncertainty (both in the ID and the MS) are taken into account. The experimental systematic uncertainties are dominated by the muon reconstruction efficiency, which results in a relative systematic uncertainty of $\sim 10\%$ ($\sim 17\%$) at $m_{\mu\mu} = 2~\TeV{}$ ($m_{\mu\mu} =~4~\TeV{}$).

\begin{table}[htb]
	\begin{center}
		\begin{tabular}{|l|c|c|c|}
			\hline
			& \textbf{Drell-Yan} & \textbf{Top} & \textbf{Diboson}\\\hline
			\textbf{Generator} & Powheg v2 & Powheg v2 & Sherpa 2.1.1\\\hline
			\textbf{Order} & NLO & NLO & NLO \\\hline
			\textbf{Shower} & Pythia 8.186 & Pythia 6.428 & Sherpa 2.1.1 \\\hline
			\textbf{PDF} & CT10 & CT10 & CT10 \\\hline
		\end{tabular}
		\caption{
		\label{tab:Backgrounds}
		Summary of background processes considered in this search.}
	\end{center}
\end{table}



\section{Results}
\begin{figure}[!h]
\begin{center}
\subfloat[][]{
\includegraphics[width=0.48\linewidth]{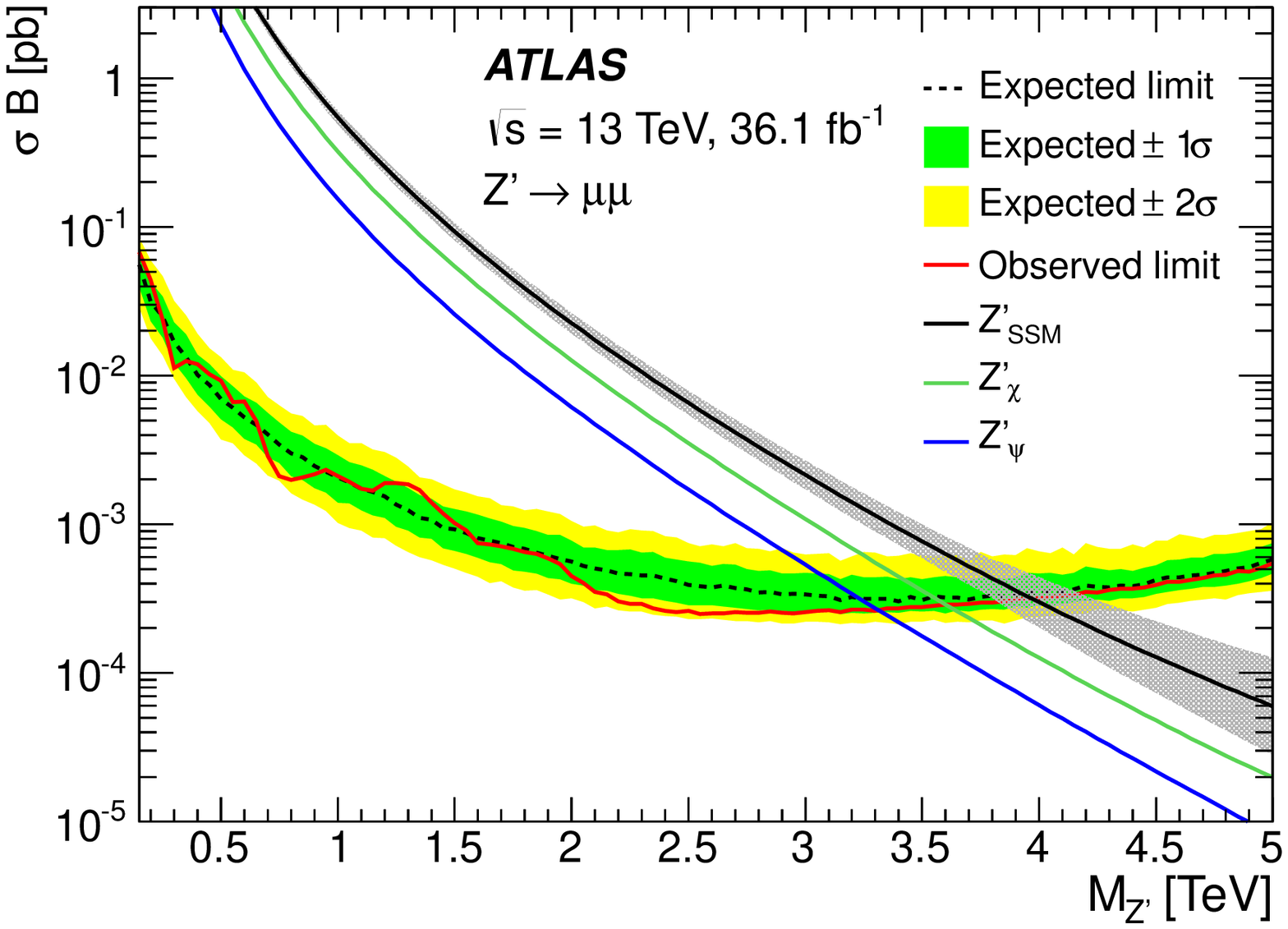}
\label{subfig:ResonantLimits}
}
\subfloat[][]{
\includegraphics[width=0.48\linewidth]{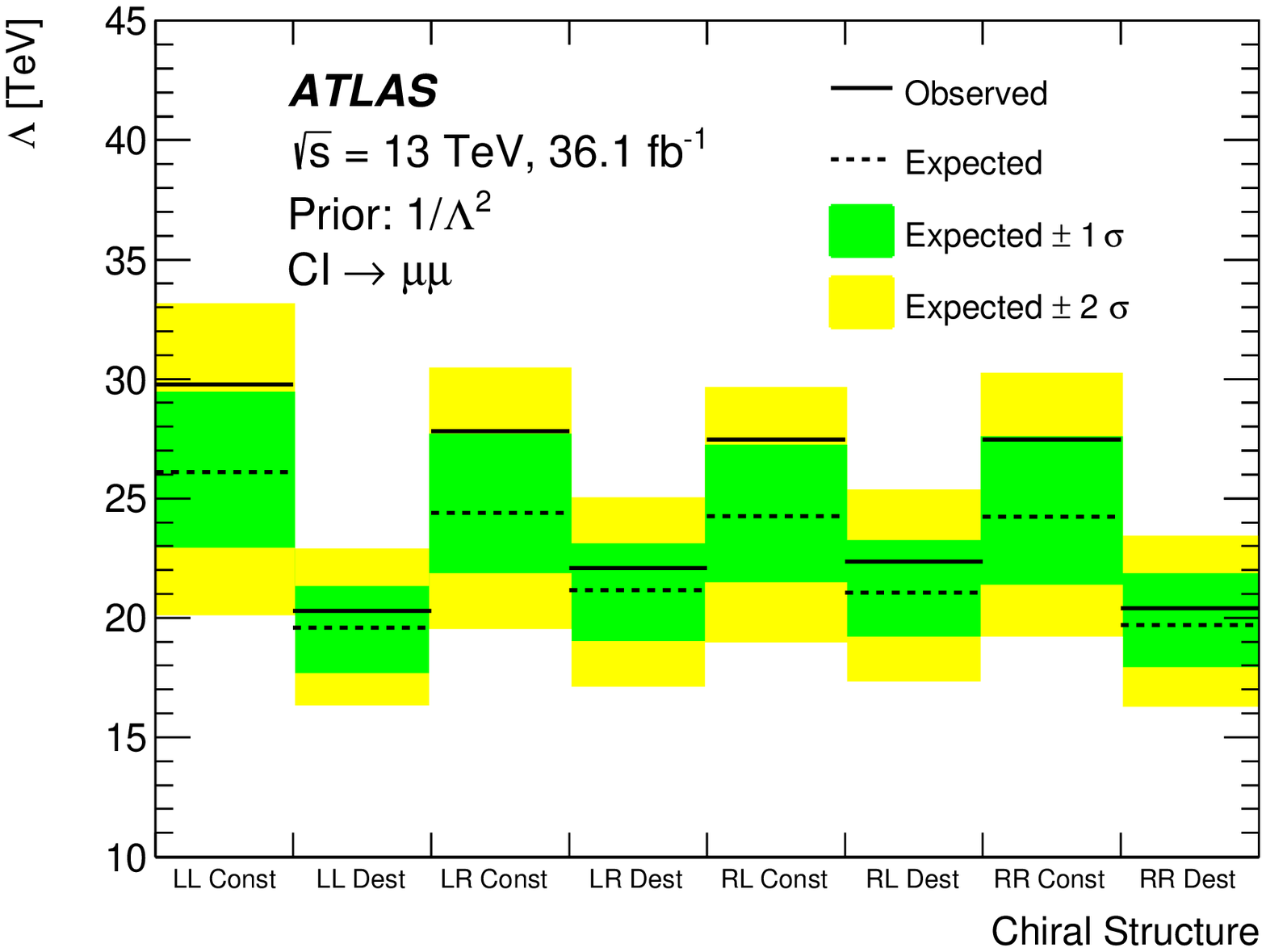}
\label{subfig:CILimits}
}
\end{center}
\caption{\protect\subref{subfig:ResonantLimits} 95\% C.L. upper limits on $\sigma B$ as a function of mass for several Z' models. \protect\subref{subfig:CILimits} 95\% C.L. lower limits on $\Lambda$, the energy scale of the Contact Interaction model with constructive (const) and destructive (dest) interference \cite{Aaboud:2017buh}.}
\label{fig:Results}
\end{figure}

In order to quantify any potential excess of data compared to background processes, we calculate the $p$-value, i.e. the probability of observing an excess at least as signal-like as the one observed in data assuming that signal is absent, as a function of mass. No significant excess was observed in the dimuon invariant mass distribution. Therefore, various theoretical models are constrained by setting limits on parameters of the models. First, upper limits on the cross-section times branching ratio ($\sigma B$) are set for various Z' models. Second, lower limits on the energy scale $\Lambda$ of Contact Interactions are set for different configurations of interference and chirality. These limits are set using a Bayesian approach \cite{Beaujean:2011zz}. Figure \ref{fig:Results} shows the various limits set for the resonant Z' models and the non-resonant Contact Interaction models. In particular, a Sequential Standard Model Z' resonance is excluded for masses below $4.0~\TeV{}$, and a Z'$_\chi$ resonance is excluded for masses below $3.6~\TeV{}$. Lower limits on the energy scale $\Lambda$ of Contact Interactions vary between $20.3~\TeV{}$ and $29.8~\TeV{}$, depending on the model.

\section{Conclusion}

A search for new high-mass phenomena decaying into muon pairs was performed using 36.1 fb$^{-1}$ of data recorded by the ATLAS detector at $\sqrt{s}=13~\TeV{}$. No significant excess above the Standard Model expectation was found, therefore lower limits at the 95\% C.L. were set on the mass of the Z' boson for various Z' models, and the energy scale of the Contact Interaction $\Lambda$ for various interference scenarios of the model.


\end{document}